\documentclass[12pt]{iopart} 

\usepackage{graphicx}
\usepackage{iopams}  
\def\mod{\hbox{mod}\,}

\begin{document}

\title{Nonadiabatic pumping in classical and quantum chaotic scatterers}

\author{A Casta\~neda$^1$, T Dittrich$^{2,3,4}$, and G Sinuco$^5$}

\address{$^1$ Max Planck Institute of Microstructure Physics,
Halle/Saale, Germany}
\address{$^3$ Departamento de F\'\i sica, Universidad Nacional de
Colombia, Bogot\'a D.C., Colombia}
\address{$^4$ CeiBA -- Complejidad, Bogot\'a D.C., Colombia}
\address{$^5$ Department of Physics and Astronomy, University of
Sussex, Brighton, UK}
\ead{$^2$ tdittrich@unal.edu.co}

\begin{abstract}
We study directed transport in periodically forced scattering
systems in the regime of fast and strong driving where the dynamics is
mixed to chaotic and adiabatic approximations do not apply. The model
employed is a square potential well undergoing lateral oscillations,
alternatively as two- or single-parameter driving. Mechanisms of
directed transport are analyzed in terms of asymmetric irregular
scattering processes. Quantizing the system in the framework of
Floquet scattering theory, we calculate directed currents on basis of
transmission and reflection probabilities obtained by numerical
wavepacket scattering. We observe classical as well as quantum
transport beyond linear response, manifest in particular in a non-zero
current for single-parameter driving where according to adiabatic
theory, it should vanish identically.
\end{abstract}

\maketitle

\section{Introduction}\label{one}

The concept of pumping in nanosystems \cite{Tho83} has emerged from
the paradigm of peristaltic pumps, simple idealized devices moving
mass and/or charge in a defined direction as two or more parameters
are cycled through a closed path \cite{AG99}. As long as the driving is
considered slow, adiabatic approximations apply. This enormously 
fruitful idea allows for a very general analytical treatment involving
quantum scattering theory and Berry phases 
\cite{Bro98,AE&00,MB02,Coh03} and has inspired a wide range of
experimental realizations, including semiconductor nanostructures, 
e.g., quantum dots driven by time-dependent gate voltages 
\cite{OK&97,BW&98} and more recently graphene sheets 
\cite{PSS09,GY&09}. Stimulated in turn by these developments,
theoretical approaches have been extended towards quantum pumps 
with fast and strong driving, leaving the adiabatic
regime---largely equivalent to linear response \cite{Coh03}---far
behind \cite{Tor05,MBE06,AS07,MB08,RG11}, as well as towards including
realistic many-body effects like dissipation and charge blocking. 

By contrast, aspects of complex dynamics involved in directed transport
in strongly driven pumps have not yet been explored in depth. From the
point of view of ballistic classical motion, peristaltic pumping amounts to
a trivial dynamics of the transported particles, moving phase-space
volumes through the device much like a viscous liquid 
\cite{AE&00,Coh03,CKS05,SC06}.
As sufficiently coherent and controllable sources of ever faster 
drivings approaching the THz regime \cite{HH&05} become available,
pumping in the realm of nonadiabatic and strongly nonlinear dynamics 
comes within reach. Scattering theory, already at the heart of quantum
pumping, provides us with an appropriate theoretical framework which 
has developed far beyond the adiabatic regime: Irregular scattering
\cite{Smi92} is a mature field with numerous applications from
celestial mechanics \cite{BTS97} down to chemical reactions
\cite{NGR86,BS88,GR89}, comprising classical as well as quantum
phenomena and approaches.

With this work, we attempt a first survey into pumps operating in the
regime of chaotic scattering and to give an overview of their
principal features. As concerns implications for directed transport of
the nonlinear dynamics induced by fast and strong forcing, there
exists a pertinent case to be followed: The concept of chaotic
ratchets \cite{Rei02}, in particular classical Hamiltonian ones
amenable to direct quantization \cite{SO&01,SDK05}, has stimulated
important insights into conditions and mechanisms of directed
transport and its quantum manifestations in this regime. Similarly,
based on a comparable body of results available on scattering at
strongly driven potentials \cite{HDR00,HDR01}, we will point out
peculiarities of pumping due to a complex dynamics, such as currents
violating linear response \cite{DGS03} and the possibility to achieve
transport with a single driven parameter
\cite{AS07,RG11,SP&11}. Conversely, the field of irregular scattering
gets enriched by considerations around symmetry breaking \cite{FYZ00}
and finds new applications with significant technological perspectives
such as the generation of polarized (pure spin) currents
\cite{DD08}. Focusing on dynamical aspects, we however abstain from
considering many-body effects in this work.

In the first, classical part of the paper, we devise a family of
elementary one-dimensional models, consisting of square potential
wells which allow for two- as well as single-parameter driving, and
analyze directed transport in these systems in terms of phase-space
structures. The quantum part briefly reviews Floquet scattering
theory as basic theoretical tool and provides numerical evidence,
obtained from wavepacket scattering, for directed currents largely
owing to the complex underlying classical motion. 
The concluding section provides an outlook to a number of questions
left open by our work.

\section{Classical chaotic pumps}\label{two}

\subsection{Models}\label{twodotone}

We seek models that (i) show irregular scattering, (ii)
exhibit directed currents yet (iii) remain simple enough to permit
an efficient numerical or even an analytical treatment, and (iv) are
amenable to experimental realizations. Square potential wells and
combinations of them almost ideally meet these conditions
\cite{HDR00,HDR01,DGS03}: They allow for different types of external
driving to induce chaotic behaviour and break symmetries as is
necessary to achieve directed transport, and they closely resemble the
potentials of semiconductor superlattices in the transverse
direction.

\begin{figure}[h!]
\centerline{\includegraphics[scale=0.95]{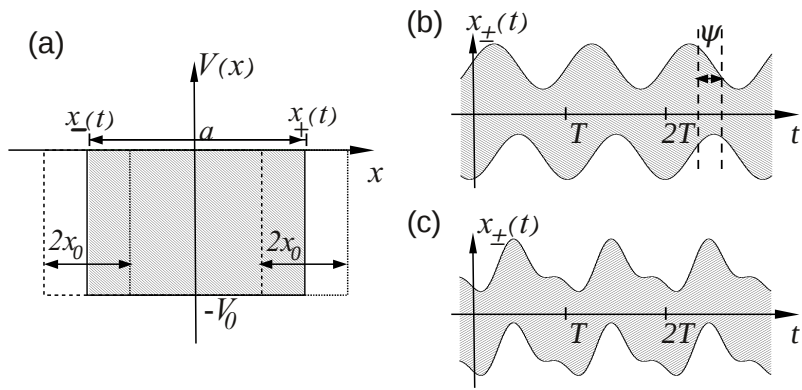}}
\caption{Models for single- and two-parameter driving. Panel a:
Laterally driven square potential well as in
eq.~(\protect\ref{losw}); b: position of the potential well
(shaded) as a function of time for two-parameter driving,
eq.~(\protect\ref{twopardriv}); c: same for single-parameter driving,
eqs.~(\protect\ref{onepardriv},\protect\ref{asymcos}).
}
\label{fig1}
\end{figure}

We consider a square well with constant depth but
lateral driving, i.e., with the positions of its walls depending on
time in some arbitrary periodic manner (fig.~1a),
\begin{equation}\label{losw}
V(x,t) = -V_0\Big[\Theta\big(x - x_{-}(t)\big) - 
\Theta\big(x - x_{+}(t)\big)\Big],
\end{equation}
where $\Theta(x)$ denotes the Heaviside step function and the wall
positions $x_\pm(t)$ are periodic functions with the same period
$T$. A lateral driving arises naturally from an AC voltage through a
Kramers-Henneberger gauge transformation \cite{Kra56,Hen58}. A phase
shift between the walls, equivalent to an oscillating width of the
well, can be motivated, e.g., by a finite propagation time across the
well of time-dependent perturbations or by a superposed gate
voltage. Mass is set $m = 1$ throughout.

We assume either two independent harmonic driving forces
(``two-parameter  driving'') with phase offset $\psi$ (fig.~1b),
\begin{equation}\label{twopardriv}
x_\pm(t) = \pm\frac{a}{2} + x_0\cos(\omega t \pm \psi/2),
\quad \omega = 2\pi/T,
\end{equation}
where $a$ is the width of the well (in all that follows, $a = 2$), or
alternatively (``single-parameter driving'') a synchronous motion of
both walls (fig.~1c) with the same periodic but otherwise arbitrary
time dependence $f(t)$,
\begin{equation}\label{onepardriv}
x_\pm(t) = \pm\frac{a}{2} + x_0 f(t), \quad f(t+T) = f(t).
\end{equation}

\subsection{Symmetries}\label{twodottwo}
In periodically driven scattering, the phase of an incoming trajectory
relative to the driving constitutes an additional scattering
parameter \cite{Smi92}, besides the incoming momentum $p_{\rm in} =
\lim_{t \to -\infty} m\dot{x}(t)$. It can be defined, e.g., as $\theta
= \omega t_{\rm in}\,{\rm mod}\,2\pi$, with $t_{\rm in}$ the arrival 
time of the scattered particle, measured by extrapolating its 
asymptotic incoming trajectory till the origin $x =
0$ \cite{DGS03}. %
As $\theta$ is typically beyond experimental control, directed
transport is considered relevant only averaged over $\theta$. In order
to avoid a systematic cancellation due to counterpropagating
trajectory pairs related by $\mathsf{P}$: ${\bf r} \to -{\bf r}$,
${\bf r} = (p,x)$, $\theta \to -\theta$ (equivalent to $\omega \to
-\omega$ or $t \to -t$ for harmonic drivings like (\ref{twopardriv})),
we have to break time-reversal invariance (TRI) of the potential. For
eq.~(\ref{twopardriv}), this is achieved if $\psi \neq 0$, $\pi$. In
eq.~(\ref{onepardriv}), it requires a function $f(t)$ 
without any symmetry or antisymmetry, $f(t_\pm-t) \neq \pm
f(t)$ for reference times $t_\pm$, which excludes in particular harmonic
forces. In the sequel, we choose (cf.\ fig.~\ref{fig1}c)
\begin{equation}\label{asymcos}
f(t) = \cos(\omega t) + \gamma \cos(2\omega t - \phi), \quad
\gamma \neq 0, \quad \phi \neq 0\,{\rm mod}\,\pi/2.
\end{equation}

For a two-parameter driving, even $\psi$ could be difficult to
control. To make sure that currents do not even vanish upon averaging
over $\psi$, we prevent the systematic pairing of trajectories related
by $\mathsf{P}$ and $\psi \to -\psi$ \cite{DGS03},  e.g., by choosing
functions for $x_\pm(t)$ not connected by any symmetry. For an
asymmetric single-parameter driving such as (\ref{asymcos}), this
problem does not arise in the first place.

A relevant issue in the context of symmetry is the energy distribution
of incoming and outgoing trajectories. In order to analyze
deterministic transport processes based on irregular scattering, it
would suffice to consider individual scattering events at a given
energy, without assuming specific distributions in the asymptotic
regions or reservoirs to keep them constant. The strongly inelastic
scattering envisaged here, however, together with symmetry breaking
suggest to have a closer look at this aspect.

\begin{figure}[h!]
\centerline{\includegraphics[scale=0.195]{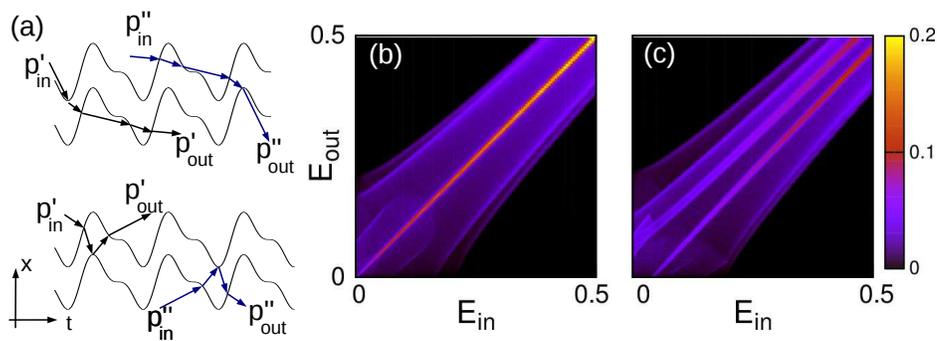}}
\caption{Symmetry and inelastic scattering. Panel a: Systematic
pairing of transmitted (above) and reflected (below) trajectories for
single-parameter driving with antisymmetric time dependence
$f(\pi/2-t) = - f(\pi/2+t)$, 
eqs.~(\protect\ref{losw},\protect\ref{onepardriv},\protect\ref{asymcos}) 
with $\phi = \pi/2$, see text and fig.~\protect\ref{fig1}c. b,c:
Probability density distributions $\langle d(E_{\rm in},E_{\rm out})
\rangle_{\theta}$ (color code) of outgoing vs.\ incoming energies for
(b) antisymmetric driving as in (a) and (c) asymmetric driving, $\phi
= \pi/4$. Other parameter values are $V_0 = 0.75$, $x_0 = 0.1$, 
$\gamma = 0.2$, $\omega = 1.0$.
}
\label{fig2}
\end{figure}

If, for example, we set $\phi = \pi/2\,\mod \pi$ in
eq.~(\ref{asymcos}) (such that $f(\pi/2\omega-t) = -
f(\pi/2\omega+t)$), trajectories come in pairs ${\bf r}''(t'') =
\mathsf{C}{\bf r}'(t')$,
$\mathsf{C}$: $(p,x,t) \to (p,-x,-t)$ (fig.~\ref{fig2}a). As a
consequence, incoming and outgoing momenta are interchanged, 
$p''_{\rm in(out)} = p'_{\rm out(in)}$, as are the energies
$E_{\rm in(out)}$. We illustrate the spread of $E_{\rm out}$ vs.\
$E_{\rm in}$ in fig.~\ref{fig2}b,c in terms of the probability density
$\langle d(E_{\rm in},E_{\rm out}) \rangle_{\theta}$. In panel (b),
the trajectory pairing is reflected in a symmetry of the distribution
with respect to the diagonal (the ridge along the diagonal corresponds
to elastic processes $E_{\rm in} = E_{\rm out}$). It is absent (c) if
the antisymmetry of the driving is broken choosing, e.g., $\phi =
\pi/4$ in eq.~(\ref{asymcos}).

Therefore, if $E_{\rm in}$ and $E_{\rm out}$ obey, say, Fermi
statistics with the same chemical potential $\mu$ and $E'_{\rm in} <
\mu < E'_{\rm out}$, the process $E''_{\rm in}(= E'_{\rm out}) \to
E''_{\rm out}(= E'_{\rm in})$ is less probable than $E'_{\rm in} \to
E'_{\rm out}$ and a symmetry-induced balance in the reflection
probabilities (fig.~\ref{fig2}a, lower panel) from either side is broken for
inelastic processes, so that they now contribute to the current. More
generally, inhomogeneities in the energy distribution, {\em even if they
occur identically on both sides}, can well enable a finite
contribution to transport of processes which otherwise would be
suppressed by some symmetry.

We show quantum results for Fermion reservoirs below
(fig.\ref{fig8}) but mainly consider transport, in classical as
well as quantum calculations, as a function of $E_{\rm in}$ without
averaging over this quantity.

\subsection{Phase-space structures}\label{twodotthree}

In the context of Hamiltonian ratchets, the optimal condition to
obtain directed currents is not hard chaos but a mixed dynamics with
chaotic and regular regions coexisting in an intricately structured
phase space \cite{SO&01,SDK05}. Pumps are no exception to this
rule. We therefore focus on systems which pertain only marginally to
the regime of chaotic scattering (defined through (i) the existence of
a chaotic repeller in phase space, (ii) self-similar deflection
functions, and (iii) exponential dwell-time distributions
\cite{Smi92}).

\begin{figure}[h!]
\centerline{\includegraphics[scale=0.975]{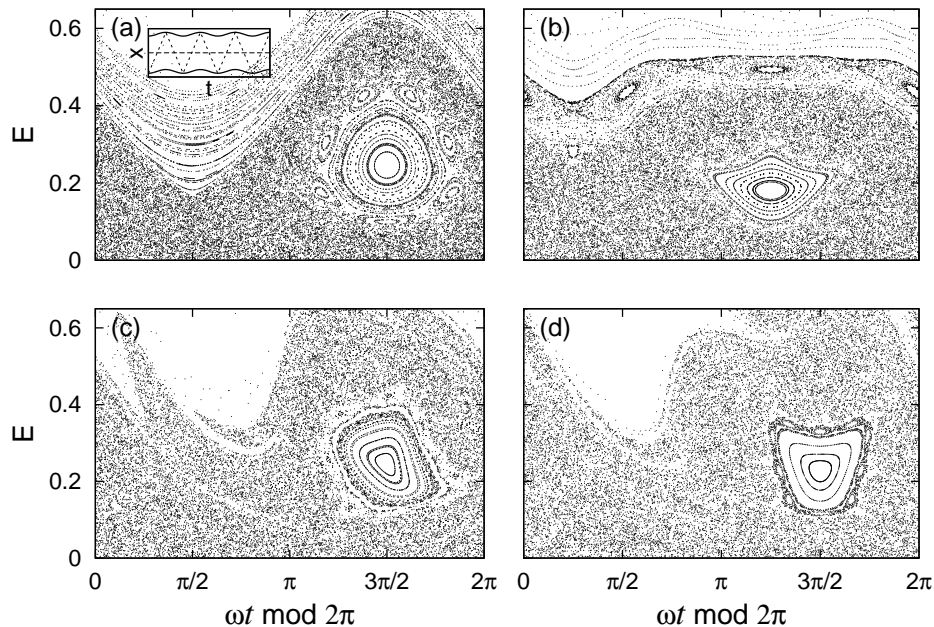}}
\caption{Poincar\'e surfaces of section (PSS, defined by $x = 0$, $p >
0$). a,b: two-parameter driving, 
eqs.~(\protect\ref{losw},\protect\ref{twopardriv}), 
c,d: single-parameter driving,
eqs.~(\protect\ref{losw},\protect\ref{onepardriv},\protect\ref{asymcos}); 
a: with TRI, harmonic driving as in
eq.~(\protect\ref{twopardriv}) with $\psi = 0$, inset: periodic orbit
underlying the prominent regular island, tracing the position $x(t)$
(short dashed) between $x_-(t)$ and $x_+(t)$ (full lines); b,c,d: TRI
broken with $\psi = \pi/2$ (b), $\gamma = 0.1$ (c,d), $\phi = 0$ (c),
$\pi/2$ (d). Other parameter values as in fig.~\protect\ref{fig2}. Large
void areas correspond to trajectories not contained in the initial
ensemble or intersecting the PSS only a few times.}
\label{fig3}
\end{figure}

In fig.~\ref{fig3} we show a set of Poincar\'e surfaces of
section, 
for two-parameter, eq.~(\ref{twopardriv}) (panels a,b) as well as for
single-parameter driving, 
eqs.~(\protect\ref{onepardriv},\protect\ref{asymcos}) 
(c,d). In all plots, we observe chaotic regions interspersed with
regular islands, characteristic of the KAM scenario
\cite{Ott93}. Typically, these islands surround periodic orbits bound
in the scattering region and not accessible from outside (see, e.g.,
the inset in panel a). The phase-space structures clearly reflect the
presence or absence of spatiotemporal symmetries of the underlying
dynamics, as is evident comparing panels (a) (TR invariant driving)
with (b),(c),(d) (TRI broken).

\subsection{Directed transport}\label{twodotfour}

Moving stepwise from phase-space structures to currents, we analyze
the relation between scattering and directed transport on two levels:
(i) locally, the outcome of individual scattering processes in terms
of asymmetries in reflection and transmission from either side, and
(ii) globally, mean currents as functions of scattering parameters
like $E_{\rm in}$. They are obtained by averaging discrete outcomes
(transmission vs.\ reflection) over, for example, the phase $\theta$
(see subs.~\ref{twodottwo}),
\begin{equation}\label{current}
\langle I\rangle_{\theta} =
\frac{1}{2\pi} \int_0^{2\pi}{\rm d}\theta\,
([T^{-+}(\theta) + R^{++}(\theta)] -
[T^{+-}(\theta) + R^{--}(\theta)]),
\end{equation}
where $T^{\pm\mp}$, $R^{\pm\pm}$, denote transmission and reflection
probabilities, resp., between the asymptotic regions $x \to
\pm\infty$. Their familiar left-right symmetry is generally broken in
the absence of a corresponding $\mathsf{P}$-invariance of the
Hamiltonian, so that, e.g., reflection from left to left can coexist
with transmission from right to left at otherwise identical parameter
values.

\begin{figure}[h!]
\centerline{\includegraphics[scale=0.975]{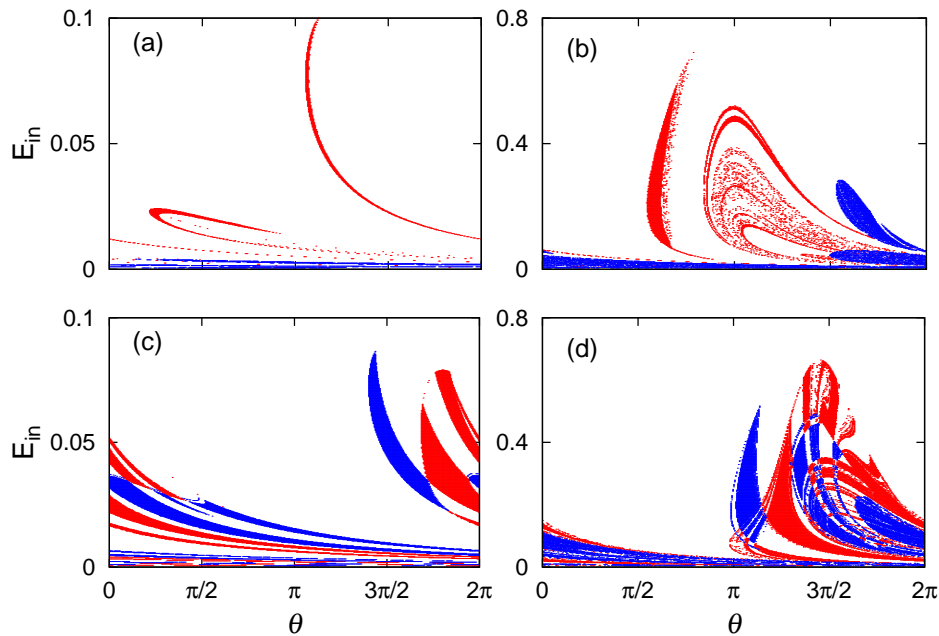}}
\caption{Directed transport in terms of individual scattering
processes in parameter space, weak (panels a,c) vs.\ strong driving
(b,d) for two-parameter,
eqs.~(\protect\ref{losw},\protect\ref{twopardriv}), (panels a,b) and
single-parameter driving,
eqs.~(\protect\ref{losw},\protect\ref{onepardriv},\protect\ref{asymcos})
(c,d). Parameter values are $V_0 = 0.75$,
$x_0 = 0.05$ (a,c), $0.4$ (b,d), $\omega = 1.0$, $\psi = \pi/2$ (a,b),
$\gamma = 0.1$, $\phi = \pi/4$ (c,d). Color code: transmission ${\rm
l} \to {\rm r}$ and reflection ${\rm r} \to {\rm r}$ blue,
transmission ${\rm r} \to {\rm l}$ and reflection ${\rm l} \to {\rm
l}$ red, else white.
}
\label{fig4}
\end{figure}

\begin{figure}[h!]
\centerline{\includegraphics[scale=0.975]{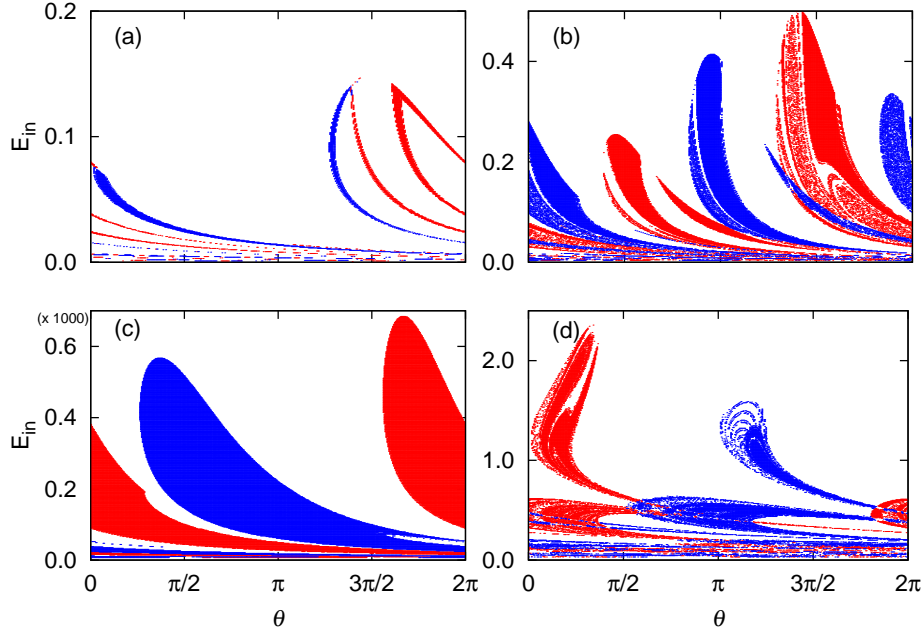}}
\caption{Directed transport in terms of individual scattering
processes in parameter space, weak vs.\ strong symmetry breaking
(panels a,b) and slow vs. fast driving (c,d) for single-parameter driving,
eqs.~(\protect\ref{losw},\protect\ref{onepardriv},\protect\ref{asymcos}).
Panel a: weakly broken symmetry, $\gamma = 0.01$, b: strongly broken
symmetry, $\gamma = 1.0$, c: low frequency, $\omega = 0.05$, d: high
frequency, $\omega = 10$. Other parameter values are $x_0 = 0.1$,
$\omega = 1.0$ (a,b), $\gamma = 0.1$ (c,d), $\phi = \pi/2$. Color code
as in fig.~\protect\ref{fig4}.
}
\label{fig5}
\end{figure}

We analyze the resulting directed transport ``locally in phase space''
in figs.~\ref{fig4}, \ref{fig5}, contrasting low with high driving
amplitude (fig.~\ref{fig4}), weak with strong symmetry breaking
(\ref{fig5}a,b). and slow with fast driving (\ref{fig5}c,d). 
Asymmetric scattering that contributes to directed currents, i.e.,
exit to the left (right), irrespective of the incoming direction, is
labeled red (blue), neutral scattering (same outcome from both sides)
white. The emerging structures reflect the relative amount of
asymmetric scattering processes in terms of their total area as well as
the complexity of the underlying phase space by forming fractals whiich 
resemble self-similar attraction basins \cite{Ott93}.

\begin{figure}[h!]
\centerline{\includegraphics[scale=0.975]{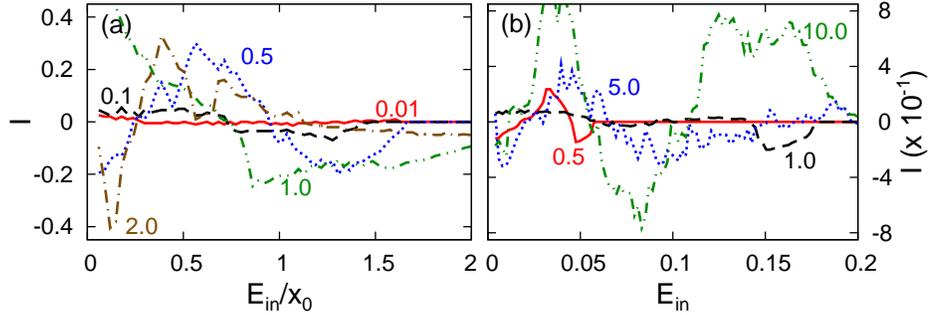}}
\caption{Average currents as functions of various
parameters. Single-parameter driving, 
eqs.~(\protect\ref{losw},\protect\ref{onepardriv},\protect\ref{asymcos}). 
Panel a: current vs.\ initial energy for different driving amplitudes,
$x_0 = 0.01$ (full red line), $0.1$ (dashed, black), $0.5$ (dotted,
blue), $1.0$ (dash-double dotted, green), $2.0$ (dash-dotted,
brown). Other parameter values are $\gamma = 0.1\times x_0$, $\phi =
\pi/2$, $\omega = 1.0$. Panel b: current vs.\ initial energy for
different driving frequencies $\omega = 0.5$ (full line, red), $1.0$
(dashed, black), $5.0$ (dotted, blue), $10.0$ (dash-double-dotted,
green). Other parameters are $V_0 = 0.75$, $x_0 = 0.1$, $\gamma =
0.1$, $\phi = \pi/2$.
}
\label{fig6}
\end{figure}

Global evidence for directed transport is provided in fig.~\ref{fig6},
comparing the dependence of the current on the incoming energy for
various values of strength (panel a) and frequency $\omega$ (b) of the
driving. As a general tendency, it increases with both parameters but
saturates as $\omega$ approaches the inverse time of ballistic flight
through the scattering region, confirming our expectation that the mixed
character of the dynamics is crucial for current generation. This is also
consistent with the fact that currents vanish towards high energy: For
$E_{\rm in} \gg V_0$, trajectories belong to the regular part of phase
space corresponding to quasi-free motion, cf.\ fig.~\ref{fig3}, hence
are always transmitted. At lower energies, current sets in as the
first trajectories are reflected for only one of the incoming directions
and reduces again as the same begins to occur also for the other
direction. Superposed on these main trends, we observe frequent sign
changes. The corresponding sensitive dependence on the parameters
\cite{DGS03,DD08} allows for a fine tuning of the current.

As is evident from figs.~\ref{fig4}, \ref{fig5}a,b, and \ref{fig6}a,
even relatively weak driving or symmetry breaking suffices to achieve
considerable transport. Conversely, this indicates that currents due to
small unintended asymmetries will be hard to avoid in a real laboratory
setup.

\section{Quantizing chaotic pumps}\label{three}

\subsection{Floquet scattering theory}\label{threedotone}

Scattering at time-dependent potentials is generally inelastic so that
standard quantum scattering theory no longer applies. If the driving
is periodic, however, large parts of that familiar setting remain
intact. A rigorous framework for the treatment of periodically
time-dependent quantum systems is provided by Floquet theory
\cite{Sam73,How79,Yaj79}, worked out for scattering systems in Refs.\
\cite{Seb93,SG94} and applied to quantum pumps in \cite{MB02}. We here
only summarize a few facts directly relevant for us:

\begin{itemize}
\item In periodically driven scattering, energy remains conserved
  ${\rm mod}\, \hbar\omega$, the photon energy associated to the
  periodic driving. For incoming and outgoing asymptotic energies this
  means
\begin{equation}\label{asymen}
  E_{\rm in/out}(\alpha,n_{\rm in/out}) = 
\epsilon_\alpha + n_{\rm in/out} \hbar\omega,
\end{equation}
where $\epsilon_\alpha$ is the \emph{quasienergy} pertaining to a
\emph{Floquet eigenstate} $\alpha$ and $n_{\rm in(out)}$ enumerate
incoming (outgoing) discrete \emph{Floquet channels}. As a
consequence, the total energy difference $E_l \equiv E_{\rm out} -
E_{\rm in} = l\hbar\omega$ is quantized, $l = n_{\rm out} - n_{\rm
in}$ counting the total number of photons exchanged with the field.

\item The time evolution can be composed as a concatenation of
  discrete steps of duration $T$ generated by the unitary Floquet
  operator
\begin{equation}\label{floqop}
\hat U_{\rm F} = 
{\cal{T}}\exp\left[-\frac{\rm i}{\hbar}
\int_0^T {\rm d}t\,\hat{H}(t)\right],
\end{equation}
where $\hat H(t)$ is the Hamiltonian and $\cal{T}$ effects time
ordering. They give rise to a \emph{discrete} dynamical group and
reduce numerical simulations of wavepacket scattering to repeated
applications of $\hat U_{\rm F}$ \cite{Seb93,SG94}. 

\item The ``on-quasienergy-shell'' scattering matrix
$S_{n_{\rm in},n_{\rm out}}(\epsilon)$ \cite{Seb93,SG94} is also
organized in terms of Floquet channels $n_{\rm in}$, $n_{\rm
out}$. For one-dimensional scattering systems, it consists of four
blocks $S^{\sigma\tau}$, $\sigma,\tau = -(+)$ denoting the left
(right) asymptotic region. 
We define partial transmission and reflection probabilities as
\begin{equation}\label{partr}
T_l^{\sigma,-\sigma}(E_{\rm in}+E_l) =
|S_{n_{\rm in},n_{\rm out}}^{\sigma,-\sigma}(E_{\rm in})|^2,\;
R_l^{\sigma,\sigma}(E_{\rm in}+E_l) =
|S_{n_{\rm in},n_{\rm out}}^{\sigma,\sigma}(E_{\rm in})|^2,
\end{equation}
and the associated total quantities as
\begin{equation}\label{tott}
  T_{\rm tot}^{\sigma,-\sigma}(E_{\rm in}) =
\sum_{l=-\infty}^\infty T_l^{\sigma,-\sigma}(E_{\rm in}),\;
  R_{\rm tot}^{\sigma,\sigma}(E_{\rm in}) =
\sum_{l=-\infty}^\infty R_l^{\sigma,\sigma}(E_{\rm in}).
\end{equation}

\item In terms of eqs.~(\ref{tott}), the pumped probability current
from left to right for fixed $E$ is obtained in analogy to the
classical probability flow (\ref{current}) as
\begin{equation}\label{qcurrent}
I(E_{\rm in}) =
T_{\rm tot}^{-+}(E_{\rm in}) + R_{\rm tot}^{++}(E_{\rm in}) -
T_{\rm tot}^{+-}(E_{\rm in}) - R_{\rm tot}^{--}(E_{\rm in}).
\end{equation}
If we assume the left and right leads to connect to
reservoirs, the charge current is given as an average weighted by
respective energy distributions $f^\sigma(E)$ \cite{MB02},
\begin{equation}\label{fcurrent}
\langle I\rangle_E = \frac{e}{\hbar} \sum_{l=-\infty}^\infty
\int_{\max(0,-E_l)}^\infty{\rm d}E\,
[T_l^{-+}(E)f^-(E) - T_l^{+-}(E)f^+(E+E_l)].
\end{equation}
For Fermi distributed electrons at zero temperature and identical
chemical potentials $\mu = E_{\rm F}$, this means
\begin{equation}\label{tzerocurrent}
\langle I\rangle_E = \frac{e}{\hbar} \sum_{l=-\infty}^\infty
\int_{\max(0,-E_l)}^{\min(E_{\rm F},E_{\rm F}-E_l)}
{\rm d}E\, [T_l^{-+}(E) - T_l^{+-}(E+E_l)].
\end{equation}

\end{itemize}

\subsection{Numerical methods}\label{threedottwo}

The matrix elements $S_{n_{\rm in},n_{\rm
out}}^{\sigma,\tau}(\epsilon)$ as basic input to further evaluation
of currents (\ref{qcurrent},\ref{fcurrent},\ref{tzerocurrent}) etc.\
are determined numerically by wavepacket scattering: An initially free
wavepacket $|\psi^\sigma_{\rm in}\rangle$ at a well-defined wavenumber
$k_0(\epsilon) = \sqrt{2m\epsilon}/\hbar$ is propagated
stroboscopically through the scattering region by repeated application
of $\hat U_{\rm F}$. Once a suitable termination criterion is
met (e.g. when the occupation probability within the interaction
region, cf.\ the integral on the r.h.s.\ of eq.(\ref{tdwell}), has
decayed to below some threshold value), the final wavepackets
$|\psi^\sigma_{\rm out}\rangle$ are decomposed into Floquet channels
at $k_l(E_{\rm in}) = \sqrt{2m(E_{\rm in}+E_l)}/\hbar$, cf.\
eq.~(\ref{asymen}), to determine the scattering matrix elements
$S_{0,l}^{\sigma,\tau}
= k_0(E_{\rm in}) \psi^\tau_{\rm out}\big(k_l(E_{\rm in})\big)/
k_l(E_{\rm in}) \psi^\sigma_{\rm in}\big(k_0(E_{\rm in})\big)$.
The Floquet operator is efficiently calculated by the $(t,t')$-method
\cite{PM&93}.

Spurious interferences between the reflected and the transmitted
wavepacket can occur if periodic boundary conditions are imposed at
the ends of the finite spatial box underlying the propagation
procedure. To prevent them, we employ absorbing boundaries
instead. Specifically, the so-called Smooth Exterior Scaling Complex
Absorbing Potentials (SES-CAPs \cite{SBM&05}), while cumbersome to
handle numerically, keep complications due to the corresponding
non-Hermitian terms in the Hamiltonian at a minimum, as they are
energy independent and leave the physical potential unscaled. 

\subsection{Quantum transport}\label{threedotthree}



\begin{figure}[h!]
\centerline{\includegraphics[scale=0.975]{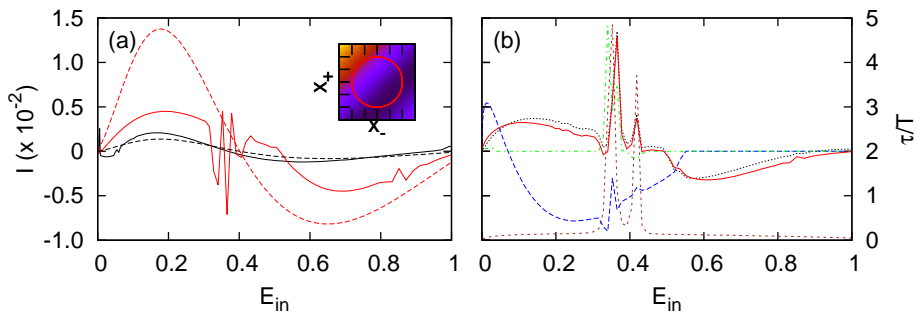}}
\caption{Quantum current as a function of the initial energy. Panel
a: Floquet approach, eq.~(\protect\ref{qcurrent}) (full lines), compared
with adiabatic theory,
eqs.~(\protect\ref{brouwer},\protect\ref{brouwermag}) (dashed), for
slow (black, $\omega = 0.1$) vs.\ fast (red, $\omega = 1.0$)
two-parameter driving,
eqs.~(\protect\ref{losw},\protect\ref{twopardriv}). Inset: parameter
cycle in $(x_-,x_+)$-space (bold red line) and fictitious magnetic
field $B(x_-,x_+)$, eq.~(\protect\ref{brouwermag}) at $E = 0.45$,
underlying the adiabatic approximation (color code: from black, $B<0$,
through red, $B=0$, through yellow, $B>0$). Panel b: Total current for
single-parameter driving,
eqs.~(\protect\ref{losw},\protect\ref{onepardriv},\protect\ref{asymcos}),
with $\omega=1.0$, $\phi=\pi/2$ (full line, red), Floquet currents in
the elastic and first and second inelastic channels, $n=0,1,2$ (dotted
black, dot-dashed green, and dashed blue lines, resp., the latter two
amplified by factors $10$ and $5\times 10^2$, resp.), and dwell time,
eq.~(\protect\ref{tdwell}) (short-dashed brown, rightmost
ordinate). Other parameters are $V_0=0.75$, $x_0=0.1$, $\hbar=0.5$.
}
\label{fig7}
\end{figure}

\begin{figure}[h!]
\centerline{\includegraphics[scale=0.975]{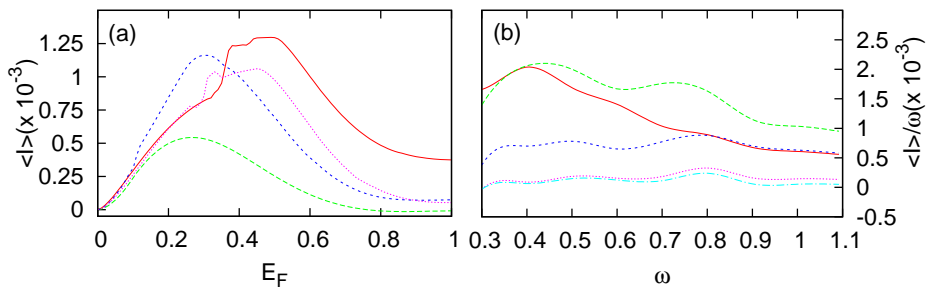}}
\caption{Fermi-averaged quantum current (\protect\ref{tzerocurrent})
for single-parameter driving, 
eqs.~(\protect\ref{losw},\protect\ref{onepardriv},\protect\ref{asymcos}).
Panel a: Mean current vs.\ Fermi energy $E_{\rm F}$ for frequencies 
$\omega = 0.1$ (full line, red), 0.3 (dashed green), 0.6 (dashed
blue), 1.0 (dotted pink). Panel b:  Scaled mean current $I/\omega$
vs.\ frequency $\omega$ at Fermi  levels $E_{\rm F} = 0.2$ (full line,
red), 0.4 (dashed green), 0.6 (dashed blue), 0.8 (dotted pink), 1.0
(dash-dotted bright blue). Other parameters are$V_0=0.75$, $x_0=0.1$,
$\hbar=0.5$.
}
\label{fig8}
\end{figure}

In order to make contact with the adiabatic approach to pumping and to
demonstrate the strongly nonadiabatic character of chaotic pumps,
we compare in fig.~\ref{fig7}a numerical results for the two-parameter
model (\ref{twopardriv}) in the Floquet approach (full lines) with the
predictions of the adiabatic scattering approach based on the Berry
phase \cite{Bro98} (broken):
For a system with a single degree of freedom, two contacts (left,
right), and two independently driven parameters $\xi_1(t)$,
$\xi_2(t)$, the probability current
through the device is
\begin{equation}\label{brouwer}
I_{\rm tot}(E) = \frac{-{\rm i}\omega}{4\pi^2} 
\int_A {\rm d}^2\xi\, B(\boldsymbol{\xi},E).
\end{equation}
The area $A$ in parameter space enclosed by the path
$\boldsymbol{\xi}(t) = (\xi_1(t),\xi_2(t))$ (cf.\ inset in
fig.~\ref{fig6}a) is penetrated by the ``magnetic field''
\begin{equation}\label{brouwermag}
B(\boldsymbol{\xi},E)\!\! = \!\!
\left[\partial_2 S_{\rm tot}^{\sigma\tau}(E) 
\partial_1 S_{\rm tot}^{\sigma\tau\dagger}(E) -
\partial_1 S_{\rm tot}^{\sigma\tau}(E)(E)
\partial_2 S_{\rm tot}^{\sigma\tau\dagger}(E)\right]^{++},
\end{equation}
with $S_{\rm tot}^{\sigma\tau}(E) = \sum_{n,n'}S_{n,n'}^{\sigma\tau}(E)$,
and $\partial_i = \partial/\partial\xi_i$.
While for slow driving (black, $\omega = 0.1$), we find appreciable
agreement, the results deviate drastically for fast driving (red, $\omega =
1.0$), indicating that strongly nonlinear transport mechanisms are
involved.

Adiabatic theory fails for single-parameter pumps, featured in
fig.~\ref{fig7}b. We decompose the total current (full line) in
Floquet channels (dashed). Although the main contribution comes from
the elastic channel ($n=0$), inelastic scattering ($n=1,2$) is also
present. 
Sharp resonances in the partial and total currents coincide largely
with peaks in the dwell time (dotted brown), measured as the fraction
of the wavepacket inside the interaction region summed over
stroboscopic time \cite{HDR01},
\begin{equation}\label{tdwell}
\tau = T \sum_{j=-\infty}^{\infty}
\int_{-a/2-x_0}^{a/2+x_0}{\rm d}x\,|\psi(x,jT)|^2.
\end{equation}
Sign changes are observed as well.
The current even persists upon Fermi averaging  (\ref{tzerocurrent})
(fig.~\ref{fig8}). As in the classical  case (fig.~\ref{fig5}b), it
increases with frequency as with energy but saturates for $\omega
\approx 1$ and $E_{\rm F} \approx 1$. Scaling the current as
$\langle I\rangle_E/\omega$ (fig.~\ref{fig8}b) reveals this systematic
deviation from the proportionality $I \sim \omega$, expected from
linear response, for $\omega \gtrsim 1$.

\section{Conclusion}\label{four}

Pumping in the non-adiabatic regime of fast and strong driving where
nonlinear classical dynamics becomes relevant, remains largely
unexplored. As a first survey into this realm, on the classical as
well as on the quantum level, we have studied directed transport
induced by irregular scattering. In particular, we have pointed out
and provided evidence for the possibility of generating currents by
driving just a single parameter, which manifestly goes beyond the
adiabatic approximation. Their sensitive parameter dependence reflects
the underlying chaotic dynamics and suggests to be harnished for
control purposes. The classical nonlinearity induces
similar nonadiabatic quantum transport: According to quantum adiabatic
transport theory \cite{Bro98}, single-parameter current generation is
impossible. Hence its occurrence for sufficiently strong and/or fast
driving cannot be explained by the mere breaking of TRI alone.
Analyzing this quantum-classical relation in the
numerically very demanding regime of small effective Planck's
constant where semiclassical methods apply \cite{RS02} remains as a
future task.

Our model has been kept utterly simple, focused on the
analysis of chaotic scattering and its impact on directed
transport. Several options are conceivable to include more
realistic details: Many-particle phenomena like charge blocking,
dissipation, decoherence, and finite-temperature energy distributions
in the reservoirs are obviously relevant. To be adequately treated,
they require sophisticated nonequilibrium methods such as the Keldysh
Green function technique 
\cite{GY&09,WWG02,AM06}. 
Other tempting perspectives to pursue are incorporating internal
degrees of freedom, such as in particular spin \cite{DD08}, as well as
pumping against a potential gradient \cite{SDK05} and the
rectification of noisy input forces.

\ack
We enjoyed illuminating discussions with Doron Cohen and Fran\c{c}ois
Leyvraz. TD acknowledges with pleasure the hospitality of Ben Gurion
University of the Negev during several stays at Beer-Sheva and
financial support by Volkswagenstiftung (grant I/$78\,235$).

\end{document}